\pdfoutput=1
\documentclass[11pt]{article}
\usepackage[margin=1in]{geometry}
\usepackage{graphicx}
\graphicspath{{figures/}}
\usepackage{booktabs}
\usepackage{tabularx}
\usepackage{longtable}
\usepackage{array}
\usepackage{authblk}

\usepackage{orcidlink}
\usepackage{xurl}
\usepackage{hyperref}
\usepackage{natbib}
\usepackage{xcolor}
\usepackage[section]{placeins}
\usepackage{float}
\usepackage{tikz}
\usetikzlibrary{arrows.meta,positioning,fit,backgrounds}

\newcommand{\dThreeN}{19,906}
\newcommand{\dThreeMean}{0.287}
\newcommand{\dThreePctHalf}{21.82}
\newcommand{\dThreePNinetyFive}{0.872}
\newcommand{\dThreeCorpusN}{99,442}
\newcommand{\dThreeSeed}{20260817}
\newcommand{\dThreeAtThree}{35.62}
\newcommand{\dThreeAtFive}{21.82}
\newcommand{\dThreeAtSeven}{12.4}
\newcommand{\dThreeSwing}{23.2}
\newcommand{\dThreeAmbiguous}{10.47}
\newcommand{\dFourNQ}{817}
\newcommand{\dFourNCat}{38}
\newcommand{\dFourTopCat}{Misconceptions}
\newcommand{\dFourTopCatN}{100}
\newcommand{\dFourMCOne}{27.7\%}
\newcommand{\dFourMCOneCI}{24.7--30.8}
\newcommand{\dFourMCOneChance}{22.6\%}
\newcommand{\dFourMCOneModel}{gpt2}
\newcommand{\dFourMCOneN}{817}

\newcommand{\bbqTotal}{58,492}
\newcommand{\bbqNCat}{11}
\newcommand{\bbqTopTwoShare}{46.4}
\newcommand{\bbqImbalance}{18.5}
\newcommand{\bbqLargest}{raceXGender}
\newcommand{\bbqSmallest}{sexualOrientation}
\newcommand{\dFiveClean}{79.3\%}
\newcommand{\dFiveCleanSD}{0.45}
\newcommand{\dFiveNSeeds}{3}
\newcommand{\dFiveNEval}{10,000}
\newcommand{\dFiveEpsEight}{2.3\%}
\newcommand{\dFiveEpsEightCI}{2.1--2.4}

\newcommand{\dFiveDrop}{77.1\%}
\newcommand{\dFiveEpochs}{15}

\newcommand{\dFiveEpsEightSeedRange}{1.4--3.1}

\newcommand{\dFiveRowFGSMZero}{79.3\% [78.9--79.8]}
\newcommand{\dFiveRowFGSMOne}{41.1\% [40.5--41.7]}
\newcommand{\dFiveRowFGSMTwo}{19.7\% [19.2--20.1]}
\newcommand{\dFiveRowFGSMFour}{5.8\% [5.5--6.0]}
\newcommand{\dFiveRowFGSMEight}{2.3\% [2.1--2.4]}
\newcommand{\dFiveRowFGSMSixteen}{2.1\% [2.0--2.3]}
\newcommand{\dFiveRowPGDZero}{79.3\% [78.9--79.8]}
\newcommand{\dFiveRowPGDOne}{34.9\% [34.4--35.4]}
\newcommand{\dFiveRowPGDTwo}{8.1\% [7.8--8.4]}
\newcommand{\dFiveRowPGDFour}{0.2\% [0.1--0.2]}
\newcommand{\dFiveRowPGDEight}{0.0\% [0.00--0.01]}
\newcommand{\dFiveRowPGDSixteen}{0.0\% [0.00--0.01]}

\newcommand{\dThreeSMSN}{5,574}
\newcommand{\dThreeSMSSpamPct}{13.4}
\newcommand{\dThreeSMSVocab}{500}
\newcommand{\dThreeSMSAP}{0.976}
\newcommand{\dThreeSMSTestN}{1,673}
\newcommand{\dThreeSMSTestSpam}{224}
\newcommand{\dThreeSMSFPThree}{15}
\newcommand{\dThreeSMSMissedThree}{18}
\newcommand{\dThreeSMSFPSeven}{2}
\newcommand{\dThreeSMSMissedSeven}{36}

\newcommand{\tierOneItems}{21}
\newcommand{\tierOneSources}{4}
\newcommand{\tierOneTeacherN}{1,987}
\newcommand{\tOnePolicyYes}{34.5\%}
\newcommand{\tOnePolicyDK}{25.8\%}
\newcommand{\tOnePolicyClear}{36.2\%}
\newcommand{\tOnePaceDK}{30.1\%}
\newcommand{\tOneCredit}{53.9\%}
\newcommand{\tTwoPersInfo}{70.9\%}
\newcommand{\tTwoLowControl}{59.3\%}
\newcommand{\tThreeScamProb}{79.5\%}
\newcommand{\tThreeScamText}{63.9\%}
\newcommand{\tThreeLostMoney}{20.8\%}
\newcommand{\tThreeKnowLot}{71.0\%}
\newcommand{\tThreeScamAI}{67.8\%}
\newcommand{\tThreeImperson}{78.0\%}
\newcommand{\tFourInaccurate}{66.4\%}
\newcommand{\tFourTrustGPT}{58.5\%}
\newcommand{\tSixConcerned}{50.9\%}
\newcommand{\tSixKTwelveNeg}{34.2\%}
\newcommand{\tSixTeachUse}{68.9\%}
\newcommand{\tSixTeachSupport}{56.6\%}
\newcommand{\tSixStudHarder}{61.9\%}
\newcommand{\tSixTeachEasier}{72.5\%}

\newcommand{\repoURL}{\url{https://github.com/ericyoc/AI-security-literacy}}
\newcommand{\platformURL}{\url{https://cyberneuralfortify.com/}}
\newcommand{\competingInterests}{The authors developed NeuralFortify
(\platformURL), which is offered to schools on a paid subscription basis. The
platform is presented in this paper as an illustrative operationalization only;
no learner data were collected, and no efficacy claim is made for it. The
empirical results in Section~\ref{sec:grounding} are produced entirely from
public and freely registered datasets by the pipeline at \repoURL\ and are
independent of the platform. The authors declare no other competing interests.}

\title{AI Security Literacy: Ecosystem-based Competency Framework and K-12 Learning Platform}
\author[1]{Eric Yocam\,\orcidlink{0000-0001-8176-3867}\thanks{Corresponding author.}}
\author[2]{Varghese Vaidyan\,\orcidlink{0000-0002-9737-242X}}
\author[1]{Paris Kalathas\,\orcidlink{0000-0002-9857-5325}}
\affil[1]{Department of Computer Science and Software Engineering, California Polytechnic State University, San Luis Obispo, CA 93407, USA}
\affil[2]{Beacom College of Computer and Cyber Sciences, Dakota State University, Madison, SD 57042, USA}
\date{}

\begin{document}
\maketitle

\begin{abstract}
Adoption of artificial intelligence in K–12 schools has outpaced guidance for
students, teachers, and administrators. Existing AI-literacy frameworks emphasize
understanding and use; cybersecurity frameworks predate AI-specific threats, and
neither defines \emph{AI security literacy} as a competency domain in its own
right. We propose an AI Security Literacy (AISL) framework of six dimensions
spanning the school ecosystem, ground its technical dimensions in public
AI-safety benchmarks and a reproducible multi-seed adversarial evaluation, and
ground its attitudinal dimensions in nationally representative survey micro-data
from U.S. teachers and adults. Only \tOnePolicyYes\ of teachers report a school
AI policy, and teachers see their own AI use as easing their work but students'
use as complicating it; on the technical side, a reported robustness figure
depends as much on the attack and the particular trained model as on the system
evaluated. NeuralFortify, a browser-based K–12 platform built on real AI models,
is presented as an illustrative operationalization.
\end{abstract}


\section{Introduction}\label{sec:intro}

Black-box artificial intelligence is now ubiquitous in students' lives, and
adoption inside schools has been correspondingly rapid. The resulting problem is
not solely a technical one about model robustness. It is also a human one:
students are frequently unsure of what they are permitted to do, teachers are
divided between embracing AI and fearing it, and many schools operate without
policy or training to guide either group \citep{rand2025guidance}.

The central claim of this paper is that \emph{AI security literacy} is best
treated as a distinct, teachable construct located on the risk, trust, and
security axis, and that its constituent phenomena can be empirically grounded
with data available today. We advance it as a proposition supported by evidence
rather than as a validated finding; discriminant validation against adjacent
literacies is future work. Existing
AI-literacy frameworks were designed to help learners understand and use AI
systems; existing cybersecurity frameworks were written before AI-specific
threats existed. Neither treats operating securely in an AI-mediated environment
as a competency domain with its own dimensions and developmental progression.

This paper makes four contributions. It defines the AISL framework across six
dimensions spanning the school ecosystem rather than just students. It grounds
the technical dimensions in public benchmarks and reproducible adversarial
evaluation, and the attitudinal dimensions in weighted secondary analysis of
nationally representative surveys of teachers and adults. It presents NeuralFortify as
an existence proof that the framework is implementable on real models. Finally,
it outlines a data-generation and validation roadmap for the field, which is
proposed rather than undertaken here. The scope is deliberately bounded: the
framework is grounded and implementable now, and validation will occur next.

\section{Background and Related Work}\label{sec:background}

This section reviews the three traditions the framework draws on: K--12
AI-literacy frameworks, adversarial machine learning, and cybersecurity
education. It then summarizes the survey evidence on AI adoption in schools that
motivates the work.

Long and Magerko's competency model \citep{long2020ai} and the AI4K12 Five Big
Ideas \citep{touretzky2019envisioning} establish the vocabulary for K–12 AI
literacy, and subsequent work extends it through a conceptual review of AI
literacy \citep{ng2021conceptualizing}, middle-school curricula
\citep{lee2021developing}, and UNESCO's competency framework for students
\citep{unesco2024students}. Across this body of work, security appears at most
under societal impact or ethics and is never an operationalized domain with its
own progression \citep{k12ailiteracy2023review, k12ai2025umbrella}.

Work on adversarial machine learning in K–12 settings is even thinner. Most
classroom AI education centers on \emph{using} or \emph{building} classifiers.
The closest adversarial effort teaches learners to \emph{break} a model through
spoofing and explainable-AI inspection \citep{breakablemachine2025}, which is
valuable but stops short of the defend-and-evaluate half of the arc. Rigorous
adversarial evaluation, meanwhile, exists almost entirely at the expert level
\citep{croce2021robustbench, athalye2018obfuscated, tramer2020adaptive}, and its
vocabulary of attacks and mitigations has been standardized for practitioners
rather than learners \citep{nist2025aml}.

The threats that such evaluation addresses are no longer confined to
laboratories. Language models can reproduce memorized training data verbatim,
exposing what a system retains \citep{carlini2021extracting}; applications built
on them can be redirected by instructions hidden in content they retrieve
\citep{greshake2023indirect}; and their fluent output can be confidently false
\citep{ji2023hallucination}. Each of these reaches ordinary users of AI systems
directly, and each later anchors a dimension of the framework proposed here.

Cybersecurity education frameworks form the third relevant tradition. GenCyber
concepts and the NICE Framework \citep{nist2020nice} predate AI-specific threats and have remained
largely siloed from AI-literacy work.

The empirical phenomenon motivating this work is documented in nationally
representative survey evidence: adoption outrunning guidance, a policy vacuum,
teacher apprehension, and student confusion about permission
\citep{rand2025guidance, rand2025districts, rand2025uneven, cdt2025embrace,
collegeboard2025criticalthinking, dec2025faculty}. One finding is particularly
consequential for framework design. In a study of university students,
policy awareness had no significant effect on behavior, whereas ethical beliefs
did \citep{studentperceptions2025}. If awareness
of the rules is not the lever, then literacy rather than policy is where
intervention must occur.

\section{The Research Gap}\label{sec:gap}

This section assesses prior work against five requirements an AI security
literacy framework must meet and names the specific gap each leaves open. It
closes by deriving the three research questions the rest of the paper answers.

Table~\ref{tab:gap} positions recent and foundational work against five
requirements that an AI security literacy framework must satisfy.

\begin{table}[H]
\centering
\caption{Recent and foundational work assessed against five requirements for an AI security literacy framework.}
\label{tab:gap}
\small
\begin{tabularx}{\textwidth}{@{}>{\raggedright\arraybackslash}Xccccc@{}}
\toprule
\textbf{Work} & \textbf{Security as a} & \textbf{Attack$\to$} & \textbf{Robustness} & \textbf{Whole} & \textbf{Empirical} \\
 & \textbf{domain} & \textbf{defend arc} & \textbf{authenticity} & \textbf{ecosystem} & \textbf{grounding} \\
\midrule
\citet{long2020ai} &   &   &   &   & $\bullet$ \\
AI4K12 \citep{touretzky2019envisioning} & $\circ$ &   &   &   & $\circ$ \\
K--12 AI literacy review \citep{k12ailiteracy2023review} & $\circ$ &   &   &   & $\bullet$ \\
Breakable Machine \citep{breakablemachine2025} & $\bullet$ & $\circ$ &   &   & $\circ$ \\
RobustBench \citep{croce2021robustbench} & $\bullet$ & $\circ$ & $\circ$ &   & $\bullet$ \\
Obfuscated gradients \citep{athalye2018obfuscated} & $\bullet$ &   & $\bullet$ &   & $\bullet$ \\
RAND / CDT surveys \citep{rand2025guidance,cdt2025embrace} & $\circ$ &   &   & $\bullet$ & $\bullet$ \\
\midrule
\textbf{AISL + NeuralFortify (this work)}          & $\bullet$ & $\bullet$ & $\bullet$ & $\bullet$ & $\bullet$ \\
\bottomrule
\multicolumn{6}{@{}l@{}}{\footnotesize $\bullet$ = addressed; $\circ$ = partially addressed; blank = not addressed.}
\end{tabularx}
\end{table}

Reading the table by column shows that each requirement is met somewhere in the
literature, but never jointly. The AI-literacy tradition in the upper rows is
strong on developmental structure and, in the review literature, on empirical
grounding, yet it treats security as a peripheral ethical concern rather than a
domain to be sequenced and assessed. The adversarial machine learning tradition
in the middle rows inverts this: it treats security as the entire subject and has
produced the field's most rigorous evaluation standards, including the
distinction between apparent and authentic robustness, but it is written for
expert practitioners and addresses neither classroom progression nor the
non-technical roles surrounding a student. The survey literature in the
penultimate row is the only body of work spanning the full school ecosystem, and
it measures the confusion this framework aims to remedy without proposing a
competency structure to remedy it.

\FloatBarrier

Table~\ref{tab:unique} makes the resulting distinction explicit by naming, for
each requirement, the specific gap in existing work and what this paper
contributes against it.

\begin{table}[H]
\centering
\caption{The specific gap in existing work against each requirement, and this paper's corresponding contribution.}
\label{tab:unique}
\small
\begin{tabularx}{\textwidth}{@{}>{\raggedright}p{0.19\textwidth}>{\raggedright}X>{\raggedright\arraybackslash}X@{}}
\toprule
\textbf{Requirement} & \textbf{Gap in existing work} & \textbf{Contribution of this work} \\
\midrule
Security as a domain &
Security appears under ethics or societal impact, without dimensions, descriptors, or progression &
Six dimensions with three-level progression, defined on the risk/trust/security axis \\
\addlinespace
Attack$\to$defend arc &
Classroom efforts teach either building or breaking; the defend and evaluate stages are absent &
A full attack$\to$extract$\to$defend$\to$evaluate sequence on real models \\
\addlinespace
Robustness authenticity &
The apparent-versus-authentic distinction exists only in expert literature &
The distinction reproduced at classroom scale and made a terminal learning objective \\
\addlinespace
Whole ecosystem &
Frameworks address students; surveys address educators; neither spans both &
Role and stakeholder treated as a cross-cutting strand across all six dimensions \\
\addlinespace
Empirical grounding &
Frameworks are asserted, then validated later if at all; no public dataset targets this construct &
Technical dimensions grounded in public benchmarks and attitudinal dimensions in national survey micro-data, before any learner study \\
\bottomrule
\end{tabularx}
\end{table}

\FloatBarrier

Stated precisely, the gap is this: no operationalized, standards-aligned
framework treats AI security as a first-class literacy domain, sequences a
complete attack-to-evaluation arc, and spans the school ecosystem. Three research
questions follow, addressed in Sections~\ref{sec:framework}
through~\ref{sec:neuralfortify} respectively. RQ1 asks what competencies
constitute AI security literacy across the school ecosystem. RQ2 asks to what
extent existing public datasets provide empirical grounding for those dimensions.
RQ3 asks how a platform built on real AI models can operationalize the framework,
and how it could in future generate data to validate it.

\section{The AI Security Literacy Framework}\label{sec:framework}

This section answers RQ1. It describes how the six dimensions were derived,
defines each with its Novice, Apprentice, and Practitioner descriptors, and
introduces the two strands that cut across all six.

\subsection{Construction method}\label{sec:construction}

The dimensions were derived through deductive synthesis rather than through empirical
clustering, which is the appropriate method when the target construct has no
existing instrument to factor-analyze. The procedure had four steps.

We began by reviewing three source traditions and extracting every competency
statement that bears on risk, trust, or security in an AI-mediated setting: the
K--12 AI-literacy frameworks \citep{long2020ai, touretzky2019envisioning}, the
cybersecurity education frameworks that schools already map against (GenCyber,
CSTA, NICE \citep{nist2020nice}, ISTE), and the adversarial machine learning literature that defines
what model attack and defense actually consist of
\citep{croce2021robustbench, athalye2018obfuscated}. This produced an initial
pool of candidate competencies, many of them near-duplicates expressed in
different vocabularies.

We then consolidated that pool under three inclusion rules. A candidate was
retained only if it was \emph{not} already covered by general AI literacy or
general cybersecurity literacy, since a dimension that either parent framework
teaches adequately does not justify a new domain. It had to be
\emph{teachable and observable} at the grades 6–8 band, which excluded several
formal-methods and governance competencies that survive in the expert
literature. Finally, it had to be \emph{instantiable against a real model or a
real policy artifact}, so that the dimension could be assessed by what a learner
does rather than by what a learner reports.

The retained competencies were mapped against the ecosystem roles. Any candidate
that changed meaning across student, teacher, and administrator was designated a
cross-cutting strand rather than a dimension, which is how role and ethics came
to sit orthogonally to the six. Consolidating the remainder yielded six
dimensions; the count is an outcome of the merge rules, not a target.

Each dimension was then checked for benchmark evidence, and the mapping in
Table~\ref{tab:datasets} is that check made explicit. A dimension for which no
public dataset or policy corpus could establish the underlying phenomenon would
have been a candidate for removal. None were removed on that basis, though the
evidence available differs sharply in kind across the six, which
Table~\ref{tab:evidence} sets out.

\FloatBarrier

The framework comprises six dimensions, summarized in Table~\ref{tab:dimensions},
each with a Novice, Apprentice, and Practitioner descriptor and each applying
across student, teacher, and leader roles.

\begin{table}[H]\centering
\caption{The six AISL dimensions and the competency each names.}
\label{tab:dimensions}
\small
\begin{tabularx}{\textwidth}{@{}llX@{}}
\toprule
& \textbf{Dimension} & \textbf{Competency} \\
\midrule
D1 & Permission \& Norms & Knowing what use is sanctioned, by whom, and how to disclose it \\
D2 & Data \& Privacy Security & Reasoning about what a system retains, exposes, or infers \\
D3 & Threat \& Manipulation Recognition & Detecting scams, phishing, and social manipulation in AI-mediated channels \\
D4 & Trust Calibration \& Verification & Matching confidence in an output to its actual reliability \\
D5 & Adversarial \& Technical Understanding & Understanding how models fail under attack and what defense does \\
D6 & Disposition \& Agency & Sustaining efficacy and appropriate skepticism rather than avoidance \\
\bottomrule
\end{tabularx}
\end{table}

The six divide along a line worth making explicit. D1, D2, and D6 concern a
learner's relationship to an AI system --- what is permitted, what is exposed,
and whether the learner engages at all --- and are largely dispositional. D3, D4,
and D5 concern the system's behavior --- how it can be manipulated, when its
output should be trusted, and how it fails under attack --- and are largely
analytic. The division is not a hierarchy, and neither half is prior to the
other: a learner who understands adversarial failure perfectly but discloses
nothing about their AI use is not securely literate, and neither is one who
follows every policy but accepts fluent output uncritically. It does, however,
explain why the two halves admit different kinds of evidence, which
Section~\ref{sec:grounding} treats at length.

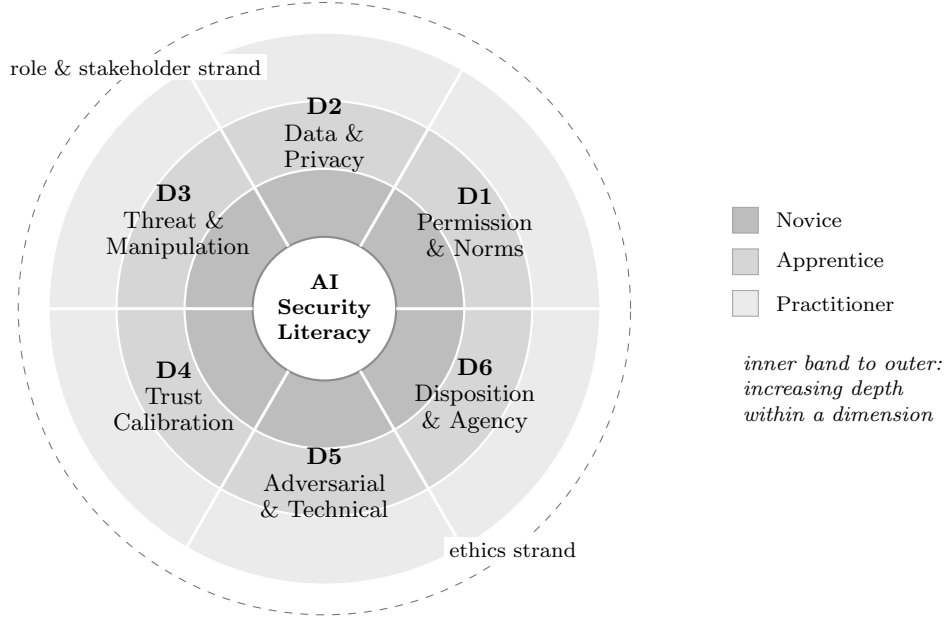
\begin{figure}[H]\centering
\begin{tikzpicture}[font=\footnotesize]
  \def\rc{0.95}\def\ra{1.85}\def\rb{2.75}\def\rd{3.65}\def\rs{4.05}
  \foreach \r/\shade in {\rd/8, \rb/16, \ra/26}
    \filldraw[fill=black!\shade, draw=white, line width=0.8pt] (0,0) circle (\r);
  \filldraw[fill=white, draw=black!45, line width=0.8pt] (0,0) circle (\rc);
  \foreach \a in {0,60,...,300}
    \draw[white, line width=1.1pt] (\a:\rc) -- (\a:\rd);
  \foreach \a/\lab in {%
      30/{\textbf{D1}\\Permission\\\& Norms},
      90/{\textbf{D2}\\Data \&\\Privacy},
     150/{\textbf{D3}\\Threat \&\\Manipulation},
     210/{\textbf{D4}\\Trust\\Calibration},
     270/{\textbf{D5}\\Adversarial\\\& Technical},
     330/{\textbf{D6}\\Disposition\\\& Agency}}
    \node[align=center, text width=18mm, font=\footnotesize\linespread{0.9}\selectfont, inner sep=0pt] at (\a:{(\rc+\rd)/2}) {\lab};
  \begin{scope}[shift={(\rs+1.35,1.0)}]
    \foreach \i/\shade/\lab in {0/26/{Novice}, 1/16/{Apprentice}, 2/8/{Practitioner}}
      {\filldraw[fill=black!\shade, draw=black!35] (0,-0.55*\i) rectangle ++(0.34,0.34);
       \node[anchor=west, font=\scriptsize] at (0.44,-0.55*\i+0.17) {\lab};}
    \node[anchor=west, font=\scriptsize\itshape, align=left, text width=30mm]
      at (0,-2.05) {inner band to outer:\\increasing depth\\within a dimension};
  \end{scope}
  \node[align=center, font=\scriptsize\bfseries] at (0,0) {AI\\Security\\Literacy};
  \draw[dashed, black!55] (0,0) circle (\rs);
  \node[fill=white, inner sep=2pt, font=\scriptsize] at (128:\rs) {role \& stakeholder strand};
  \node[fill=white, inner sep=2pt, font=\scriptsize] at (308:\rs) {ethics strand};
\end{tikzpicture}
\caption{The six AISL dimensions as sectors, the three progression levels as
concentric bands, and the two cross-cutting strands as an outer ring.}
\label{fig:framework}
\end{figure}

Figure~\ref{fig:framework} shows the resulting structure. The sectors are
adjacent rather than ordered: no dimension precedes another, and a learner
advances outward within each one at its own rate. The strands are drawn as a
ring because they modify every sector rather than occupying one.

Table~\ref{tab:progression} gives the three-level progression for each dimension.

\begin{table}[H]\centering
\caption{Novice, Apprentice, and Practitioner descriptors for each AISL dimension.}
\label{tab:progression}
\scriptsize
\begin{tabularx}{\textwidth}{@{}lXXX@{}}
\toprule
& \textbf{Novice} & \textbf{Apprentice} & \textbf{Practitioner} \\
\midrule
\textbf{D1} &
Identifies whether a given AI use requires disclosure &
Interprets a contextual acceptable-use policy and applies it to routine cases &
Resolves ambiguous cases and justifies a disclosure decision to others \\
\addlinespace
\textbf{D2} &
Recognizes that a system retains what is submitted to it &
Distinguishes what a system stores, exposes, and infers &
Chooses what to withhold, and can say what an inference would reveal \\
\addlinespace
\textbf{D3} &
Recognizes common manipulation cues in a message &
Judges a message against a detector's verdict and explains disagreement &
Sets an operating threshold and defends the resulting error trade-off \\
\addlinespace
\textbf{D4} &
Recognizes that a fluent answer can be wrong &
Verifies an output against an independent source before acting &
Calibrates confidence to evidence, and states what would change the verdict \\
\addlinespace
\textbf{D5} &
Recognizes that a small input change can break a model &
Executes an attack and applies a defense, reading the confidence change &
Evaluates a defense adaptively, separates genuine from apparent robustness, and reports a robustness figure with the attack and evaluation that produced it \\
\addlinespace
\textbf{D6} &
Engages with an AI system rather than avoiding it &
Sustains appropriate skepticism without disengaging &
Acts on judgment under uncertainty and helps others do the same \\
\bottomrule
\end{tabularx}
\end{table}

The progression is defined by what a learner can do unaided, not by content
coverage. The Practitioner descriptors deliberately require a defensible
judgment rather than a correct answer, because every dimension in this framework
terminates in a decision under uncertainty: which threshold, which disclosure,
which evidence is sufficient. That is also what makes the levels assessable from
behavior rather than from self-report.

\FloatBarrier

Two strands cut across all six dimensions rather than sitting within any one of
them. The role and stakeholder strand recognizes that the same competency looks
different for a student, a teacher, and an administrator: permission means
requesting for one and granting for another. The ethics strand recognizes that
every technical capability in this framework is dual-use, and that teaching
learners to attack a model obliges the curriculum to address why and when they
should not.

The dimensions are concurrent rather than sequential. A learner does not complete
D1 before beginning D3; each dimension advances at its own rate, and the
progression levels describe depth within a dimension rather than stages through
the framework as a whole.

\section{Empirical Grounding Using Public Datasets}\label{sec:grounding}

This section answers RQ2 by testing whether public data establish the
phenomena each dimension names. Tier 1 draws on national survey micro-data from
teachers and adults, and Tier 2 on public AI benchmarks and a multi-seed
adversarial evaluation.

The entire analysis is reproducible from the accompanying notebook. Every artifact,
including dataset caches, model checkpoints, figures, and results tables, is
written to persistent storage, and each stage records its completion in a status
file so that a rerun resumes rather than repeats. Numerical values quoted in this
section are read directly from the pipeline's generated macro file rather than
being transcribed by hand; the only exceptions are published figures cited from
their sources.

\FloatBarrier

The grounding strategy is organized into two tiers, shown in
Table~\ref{tab:datasets}. Tier 2 establishes that the technical dimensions name
real, measurable phenomena and run entirely on public data. Tier 1 maps items
from nationally representative surveys of teachers and adults to the dimensions,
using micro-data available under free registration.

\begin{table}[H]\centering
\caption{Datasets grounding each dimension, what each establishes, and how it is accessed.}
\label{tab:datasets}
\scriptsize
\begin{tabularx}{\textwidth}{@{}>{\raggedright}p{0.27\textwidth}>{\raggedright}p{0.11\textwidth}>{\raggedright}X>{\raggedright\arraybackslash}p{0.15\textwidth}@{}}
\toprule
\textbf{Tier / Dataset} & \textbf{Dim.} & \textbf{What it establishes} & \textbf{Access} \\
\midrule
RAND ATP, Fall 2025 \citep{rand2025gallupatp} & D1, D6 & Teacher-reported school AI policy, policy clarity, AI use, and attitudes & Free registration \\
Pew ATP Wave 142 \citep{pew2024w142} & D1, D4 & Norms of AI attribution; trust in AI-sourced information & Free registration \\
Pew ATP Wave 152 \citep{pew2024w152} & D2, D3, D4, D6 & Concern about privacy, impersonation, and inaccuracy; perceived control; disposition & Free registration \\
Pew ATP Wave 168 \citep{pew2025w168} & D3 & Scam exposure, losses, and self-assessed protection & Free registration \\
SMS Spam Collection \citep{almeida2011sms} & D3 & Scam-detection labels are threshold-dependent & CC BY 4.0 \\
RealToxicityPrompts \citep{gehman2020realtoxicity} & D3 & Harm is unsignaled by visible text; labels are threshold-dependent & Apache-2.0 \\
TruthfulQA \citep{lin2021truthfulqa} & D4 & Misconception susceptibility is enumerable and measurable & Apache-2.0 \\
BBQ \citep{parrish2022bbq} & D3, D4 & Bias is enumerable by social attribute & Open \\
CIFAR-10 \citep{krizhevsky2009cifar} + FGSM/PGD (this work) & D5 & Attack success is a property of the attack and of the model & Public \\
\bottomrule
\end{tabularx}
\end{table}

\FloatBarrier

Because the evidence available for each dimension differs in kind and not merely
in quantity, Table~\ref{tab:evidence} states the class of support each dimension
currently has. It should be read as a map of what this paper does and does not
establish.

\begin{table}[H]\centering
\caption{Class of evidence currently supporting each dimension.}
\label{tab:evidence}
\small
\begin{tabularx}{\textwidth}{@{}>{\raggedright}p{0.27\textwidth}>{\raggedright}p{0.20\textwidth}>{\raggedright\arraybackslash}X@{}}
\toprule
\textbf{Evidence class} & \textbf{Dimensions} & \textbf{What it does and does not establish} \\
\midrule
Population survey evidence &
D1, D2, D3, D4, D6 &
Weighted national estimates show that the named phenomena are prevalent among
teachers and adults; single items from separate samples, so no measurement
properties and no learner data \\
\addlinespace
Technical phenomenon evidence &
D3, D4, D5 &
The named phenomenon is shown to be real and measurable on public data; does not
establish that the dimension is a distinct latent factor in learners \\
\addlinespace
Future construct-validation evidence &
All six &
Dimensional structure, reliability, and discriminant validity against adjacent
literacies; not undertaken here (Section~\ref{sec:future}) \\
\bottomrule
\end{tabularx}
\end{table}

No dimension in this paper carries construct-validation evidence. D5 rests on
technical evidence alone and D1, D2, and D6 on survey evidence alone, while D3
and D4 carry both; that difference in kind, together with the absence of learner
data, is why the framework is offered as grounded rather than validated.

\FloatBarrier

\subsection{Tier 1: construct relevance from teacher and adult survey micro-data}

Tier 1 draws on four nationally representative survey files obtained under free
registration: the Fall 2025 RAND American Teacher Panel survey on teacher careers
and AI, fielded with Gallup \citep{rand2025gallupatp}, and three waves of the Pew
Research Center American Trends Panel, a probability-based panel of U.S. adults
\citep{pew2024w142, pew2024w152, pew2025w168}. Items were assigned to dimensions
from their questionnaire wording, and the response categories counted for each
item are fixed in the pipeline and printed beside every estimate. Each estimate
is the weighted share of respondents giving the stated response, excluding item
non-response, with a 95\% interval computed from the Kish effective sample size.
In all, the analysis yields \tierOneItems\ estimates from \tierOneSources\
sources, and the teacher estimates rest on up to $n={}$\tierOneTeacherN\ public
school teachers.

\begin{figure}[H]\centering
\includegraphics[width=0.95\textwidth,height=0.72\textheight,keepaspectratio]{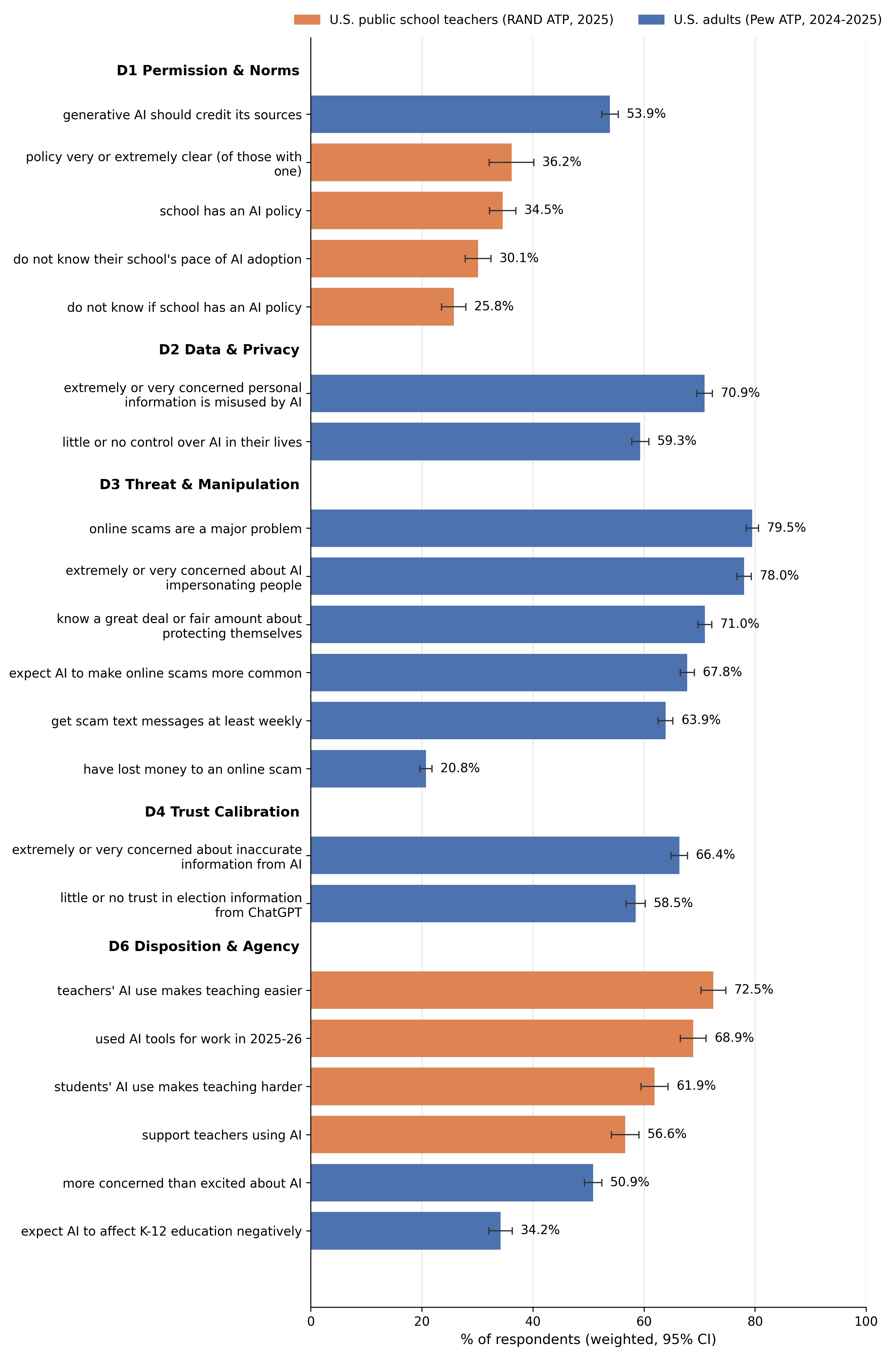}
\caption{Weighted share of respondents giving the
stated response for each mapped survey item, grouped by dimension, with 95\%
intervals.}\label{fig:tier1}
\end{figure}

The permission problem D1 names is visible among teachers directly. Only
\tOnePolicyYes\ of teachers report that their school has a policy on how AI may
be used, and \tOnePolicyDK\ do not know whether one exists; among those whose
school has a policy, only \tOnePolicyClear\ describe it as very or extremely
clear, and \tOnePaceDK\ cannot say whether their school is adopting AI too quickly
or too slowly. Most teachers are therefore working without a rule they can state.
Norms are unsettled among adults as well, of whom \tOneCredit\ hold that
generative AI should credit the sources it draws on, and among students: in
Pew's 2024 survey of teens, 54\% judged ChatGPT acceptable for researching new
topics but only 18\% for writing essays \citep{pew2025teenschatgpt}. Permission,
in other words, is task-specific and widely uncertain, which is why D1 is framed
as resolving cases rather than memorizing a policy.

For D2, \tTwoPersInfo\ of adults are extremely or very concerned that AI will
lead to people's personal information being misused, and \tTwoLowControl\ feel
they have little or no control over whether AI is used in their lives. Concern
without control is the condition D2 addresses: the competency it names is
reasoning about what a system retains, exposes, and infers, and so deciding what
to withhold, which is the agency these respondents report lacking.

For D3, \tThreeScamProb\ of adults describe online scams as a major problem,
\tThreeScamText\ receive scam text messages at least weekly, and \tThreeLostMoney\
have lost money to an online scam. Self-assessed protection is high, with
\tThreeKnowLot\ saying they know a great deal or a fair amount about protecting
themselves, so reported knowledge alone is not evidence of protection. Adults also
expect the threat to grow with AI: \tThreeScamAI\ expect AI to make online scams
more common, and \tThreeImperson\ are extremely or very concerned about AI being
used to impersonate people.

For D4, \tFourInaccurate\ of adults are extremely or very concerned about people
getting inaccurate information from AI, and among adults asked about ChatGPT,
\tFourTrustGPT\ report little or no trust in its information about the 2024
presidential election. That pattern is consistent with skepticism on a
high-stakes topic but is not evidence of calibration, since trust is reported
rather than set against accuracy; the measured failure rate in the Tier 2 D4
analysis supplies the other half of that comparison.

For D6, adults lean toward concern: \tSixConcerned\ are more concerned than
excited about the increasing use of AI, and \tSixKTwelveNeg\ expect its effect on
K--12 education to be negative. Teachers, by contrast, are already users:
\tSixTeachUse\ used AI tools in their work in the 2025--26 school year, and
\tSixTeachSupport\ support teachers using AI. The sharpest result is an asymmetry
of role. \tSixTeachEasier\ of teachers expect teachers' own use of AI to make
teaching easier, while \tSixStudHarder\ expect students' use to make it harder.
Teachers regard AI as a tool in their own hands and a problem in their
students', which is the gap the role strand of the framework exists to close:
the same competency, exercised by the learner and not only by the teacher.

This evidence has three boundaries. The Pew respondents are adults, not
students, and the teacher items concern teachers' own use and their schools'
policies; neither measures a learner's competency. Each dimension rests on a few
single items drawn from separate cross-sectional samples, so no reliability or
factor structure can be estimated, and none is claimed. And the
item-to-dimension assignment is a judgment of face validity made from the
questionnaires, not an empirically derived mapping. Tier 1 therefore establishes
that the phenomena behind D1, D2, D3, D4, and D6 are prevalent and measurable in
the national population surrounding a student, which is the construct relevance
this tier was designed to supply.

\FloatBarrier

\subsection{Tier 2: phenomenon-level construct relevance}

Tier 2 asks what public data can establish about the three technical dimensions.
The evidence it yields is \emph{phenomenon-level construct relevance}: it shows
that the phenomena each dimension names are real, measurable, and behave in ways
a learner must understand. It does not establish that D3, D4, and D5 form
distinct latent dimensions in learners, which is a psychometric question no
public benchmark can answer and which Section~\ref{sec:future} assigns to a
validation study. We use the narrower term throughout and never claim that Tier 2
validates the dimensional structure. All Tier 2 analyses use a fixed seed
(\dThreeSeed); the adversarial evaluation additionally replicates across
\dFiveNSeeds\ independently trained models.

\subsubsection*{D3: threat and manipulation recognition}

RealToxicityPrompts is constructed by sampling sentences uniformly from four
equal-width toxicity bands, then splitting each sentence into a prompt and a
continuation and re-scoring both halves independently
\citep{gehman2020realtoxicity}. That construction is what makes the corpus
suitable for D3: it permits direct comparison between the toxicity of a whole
sentence and the toxicity of the fragment a reader encounters first. From a
seeded random sample of the \dThreeCorpusN-prompt corpus, $n={}$\dThreeN\ prompts
carried non-null Perspective toxicity scores, with mean \dThreeMean\ and 95th
percentile \dThreePNinetyFive.

\begin{figure}[H]\centering
\includegraphics[width=0.95\textwidth,height=0.72\textheight,keepaspectratio]{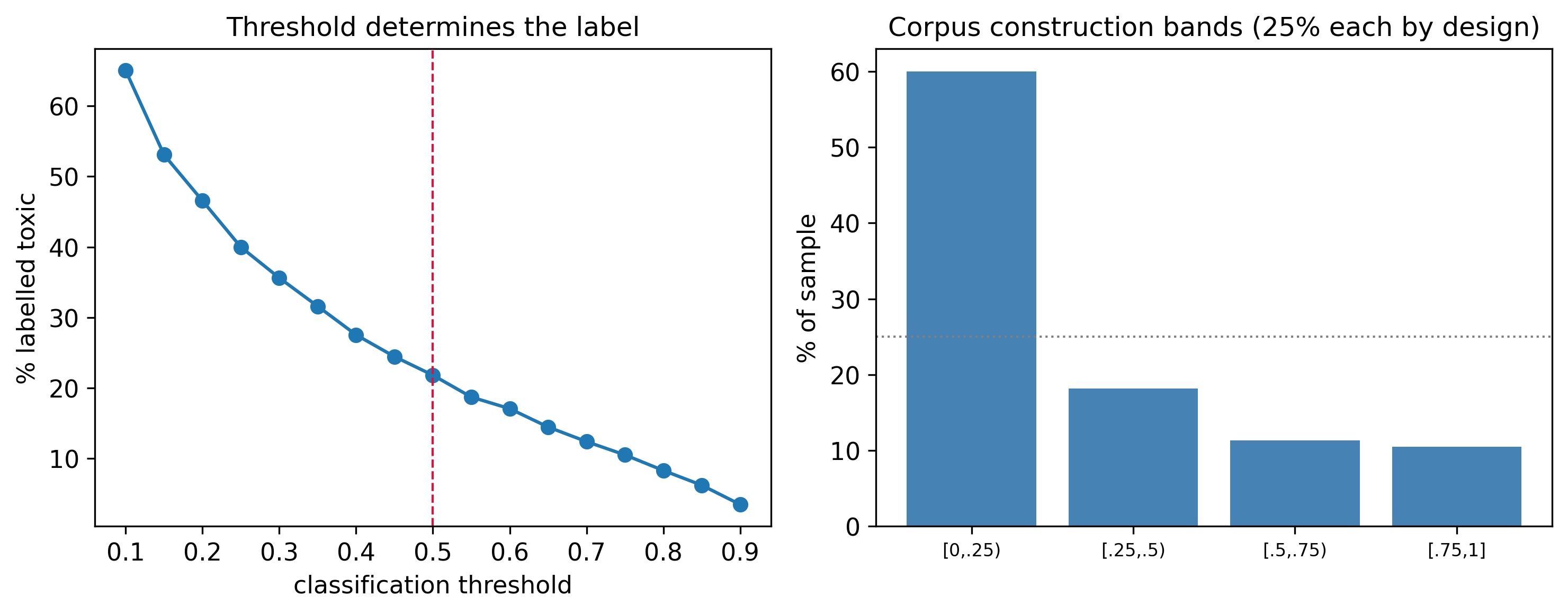}
\caption{Share of prompts labeled toxic as a
function of threshold (left), and prompt-half toxicity across the four bands from
which parent sentences were sampled equally (right).}\label{fig:d3threshold}
\end{figure}

The right panel of Figure~\ref{fig:d3threshold} carries the more consequential
result. Although the underlying sentences were drawn in equal proportions from all
four toxicity bands, the prompt halves concentrate heavily in the lowest band,
far from the 25\% design target marked on the plot. Within this corpus, an
initial text fragment can therefore substantially underrepresent the harmfulness
of the complete passage. That is the recognition problem D3 targets, demonstrated
on a corpus deliberately built to contain harmful material in equal measure
rather than inferred from an estimated prevalence. We claim it for this corpus
and this construction procedure, not as a general property of AI-mediated
communication.

The left panel establishes a second and independent D3 claim. The toxic label is
not a property of text but a consequence of a chosen operating point. Raising the
threshold from $0.3$ to $0.7$ moves the share of prompts labeled toxic from
\dThreeAtThree\% to \dThreeAtSeven\%, a swing of \dThreeSwing\ percentage points,
with \dThreeAmbiguous\% of prompts falling in the ambiguous $0.4$--$0.6$ band
around the conventional cutoff, where the share is \dThreeAtFive\%. A learner who
treats a moderation verdict as a fact about content, rather than as a decision
made at a threshold someone selected, has misunderstood the mechanism.

The same mechanism governs the scam detection that D3 most directly concerns. On
the SMS Spam Collection \citep{almeida2011sms}, \dThreeSMSN\ real text messages
of which \dThreeSMSSpamPct\% are spam, a multi-hot bag-of-words classifier over a
\dThreeSMSVocab-word vocabulary, the same input representation as the platform's
own scam model, reaches an average precision of \dThreeSMSAP\ on a held-out set
of \dThreeSMSTestN\ messages containing \dThreeSMSTestSpam\ scams. The detector
is strong, and its operating point is still a choice. At a threshold of $0.3$ it
raises \dThreeSMSFPThree\ false alarms and misses \dThreeSMSMissedThree\ scams; at
$0.7$ it raises \dThreeSMSFPSeven\ false alarms but misses \dThreeSMSMissedSeven.
No threshold removes both errors, and choosing between them is the judgment the
D3 Practitioner descriptor asks a learner to make and defend.

\begin{figure}[H]\centering
\includegraphics[width=0.95\textwidth,height=0.72\textheight,keepaspectratio]{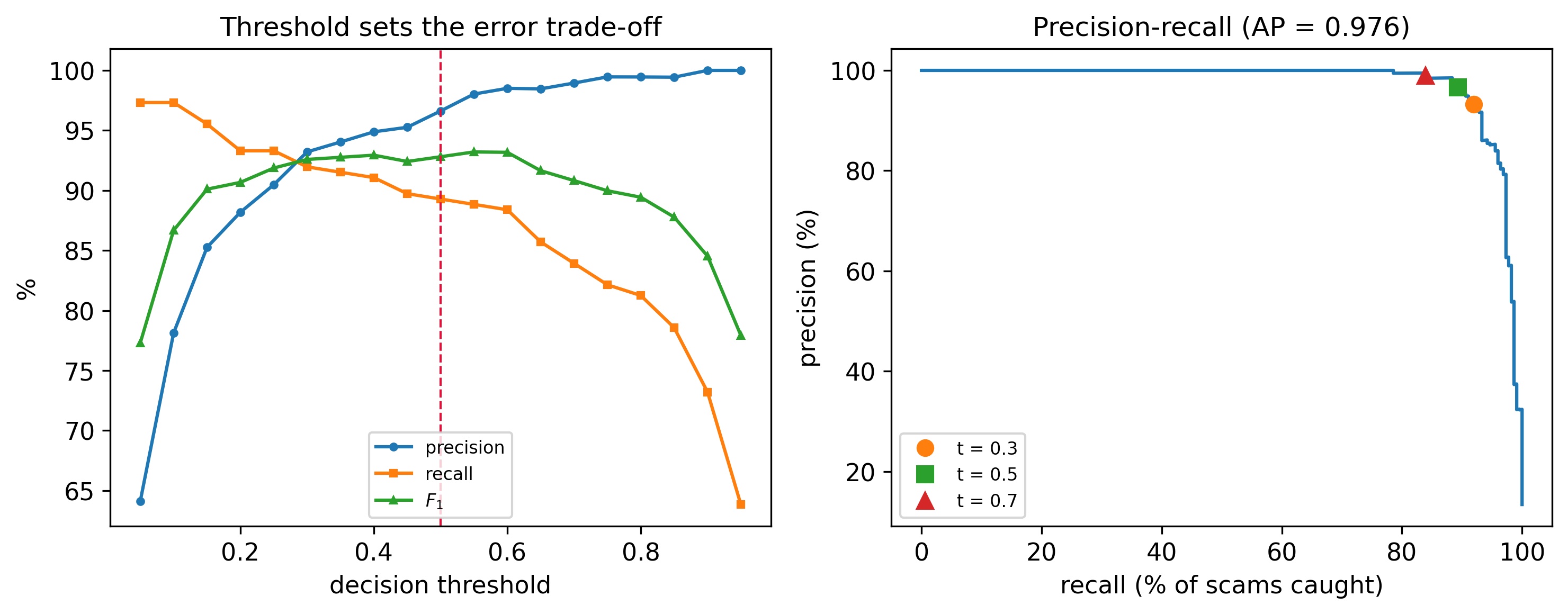}
\caption{Precision, recall, and $F_1$ of a scam-text
classifier as a function of decision threshold (left), and its precision--recall
curve with three operating points marked (right).}\label{fig:d3sms}
\end{figure}

\begin{figure}[H]\centering
\includegraphics[width=0.75\textwidth,height=0.72\textheight,keepaspectratio]{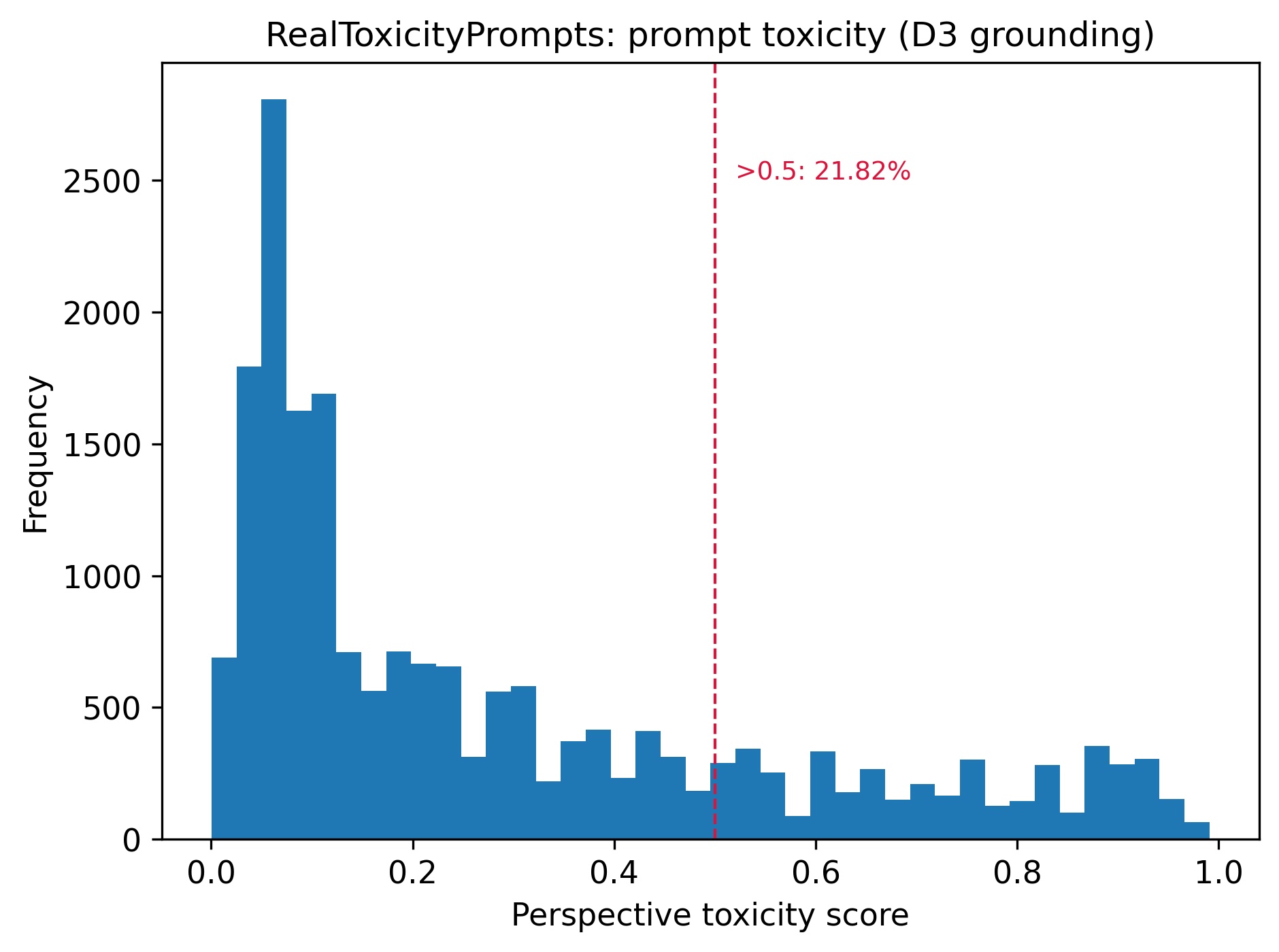}
\caption{Distribution of prompt-level Perspective toxicity
scores across the seeded sample.}\label{fig:d3}
\end{figure}

Figure~\ref{fig:d3} shows the underlying distribution from which both RealToxicityPrompts results are
drawn, with \dThreePctHalf\% of prompts above the conventional $0.5$ threshold.
The long right tail is the population the threshold analysis partitions, and the
concentration at the low end is the population the band analysis explains.

\FloatBarrier

\subsubsection*{D4: trust calibration}

The generation configuration of TruthfulQA comprises \dFourNQ\ questions across
\dFourNCat\ categories, each written so that a plausible-sounding but false
answer is the tempting one; the largest category is \dFourTopCat, with
\dFourTopCatN\ items.

\begin{figure}[H]\centering
\includegraphics[width=0.7\textwidth,height=0.72\textheight,keepaspectratio]{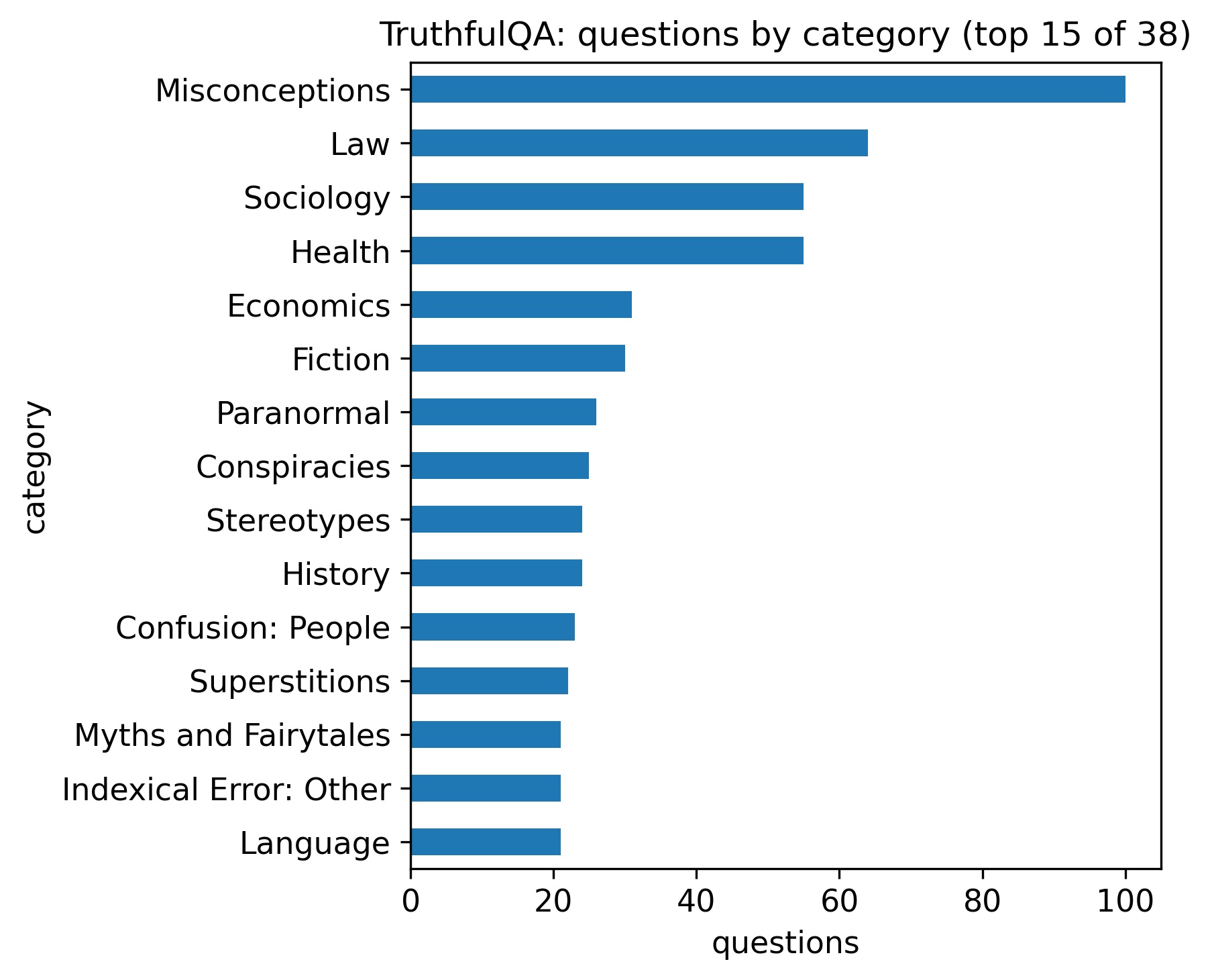}
\caption{TruthfulQA questions by category, showing
the fifteen largest of \dFourNCat\ categories.}\label{fig:d4}
\end{figure}

The breadth visible in Figure~\ref{fig:d4}, spanning health and law through
conspiracies and common misconceptions, establishes the relevance of the
verification and trust-calibration competencies D4 names. Misplaced confidence in
fluent output is a general and enumerable failure mode rather than a niche one,
which is why trust calibration can be taught as a transferable skill rather than
as a list of topics to distrust.

Category counts establish coverage, however, not difficulty, so we additionally
report an observed failure rate. Scoring \dFourMCOneModel\ \citep{radford2019gpt2} on the
multiple-choice configuration by mean per-token log-likelihood, the model selects
the true answer on \dFourMCOne\ of \dFourMCOneN\ questions (95\% CI
\dFourMCOneCI), against a chance level of \dFourMCOneChance\ given a mean of five
options per item.

\begin{figure}[H]\centering
\includegraphics[width=0.7\textwidth,height=0.72\textheight,keepaspectratio]{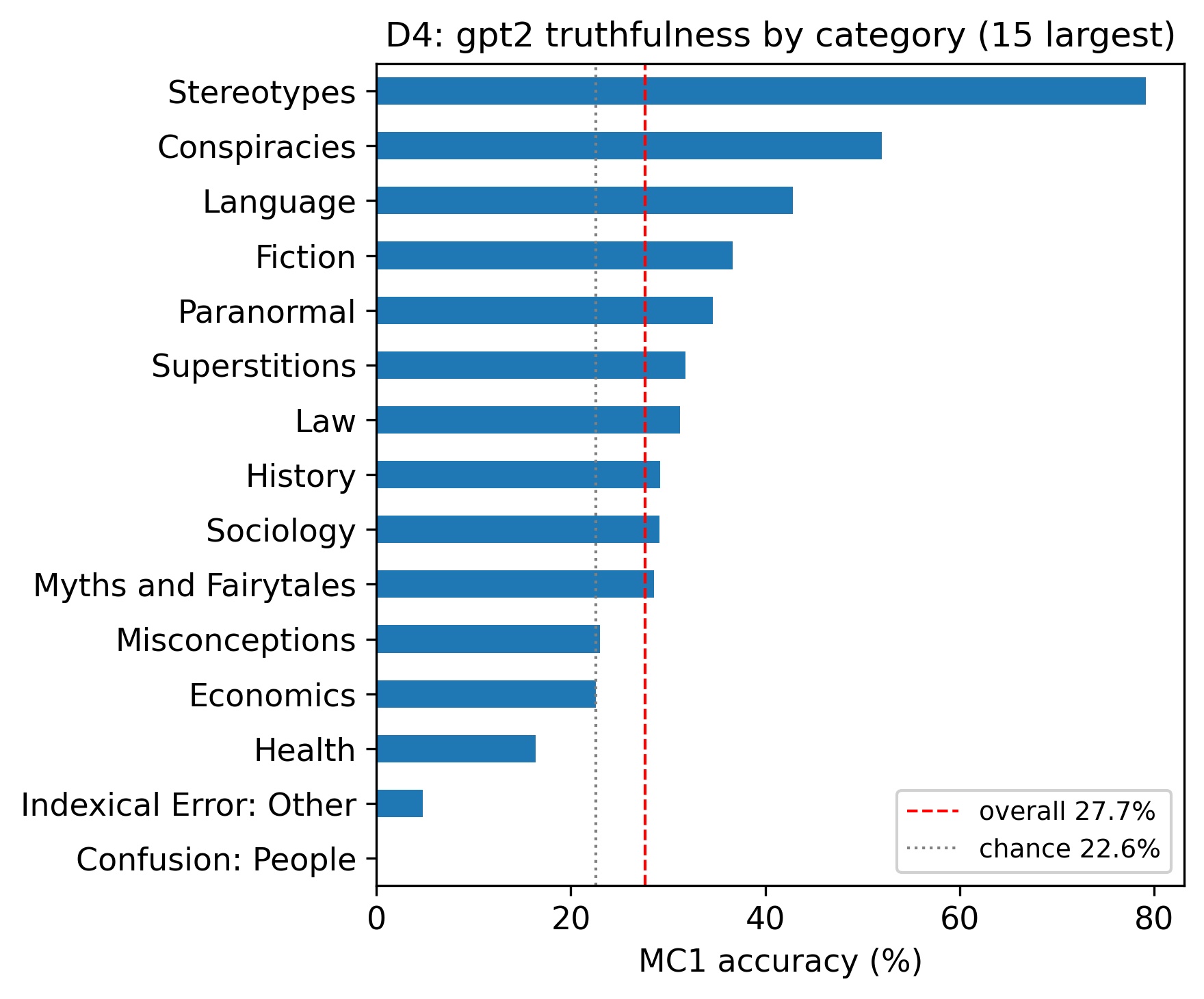}
\caption{Truthfulness of \dFourMCOneModel\ by
category for the fifteen largest categories, against the overall rate and the
chance level.}\label{fig:d4baseline}
\end{figure}

Figure~\ref{fig:d4baseline} shows how thin that margin is and how unevenly it is
distributed: the interval clears chance, but only just, and several categories
fall below it. A model that produces fluent prose is therefore barely better than
guessing at distinguishing a true answer from a plausible false one on this
benchmark, which is precisely the gap trust calibration exists to close.
Fluency at production does not confer reliability at selection, and a learner who
reads the former as evidence of the latter has made the error D4 names. The
benchmark supplies the failure \emph{opportunities} and now a measured failure
\emph{rate}; it does not itself measure calibration, which requires subjective
confidence set against objective accuracy and is a property of a learner or a
model rather than of a corpus.

\FloatBarrier

\subsubsection*{D3 and D4: social bias}

BBQ contributes \bbqTotal\ items across \bbqNCat\ social-attribute categories.

\begin{figure}[H]\centering
\includegraphics[width=0.7\textwidth,height=0.72\textheight,keepaspectratio]{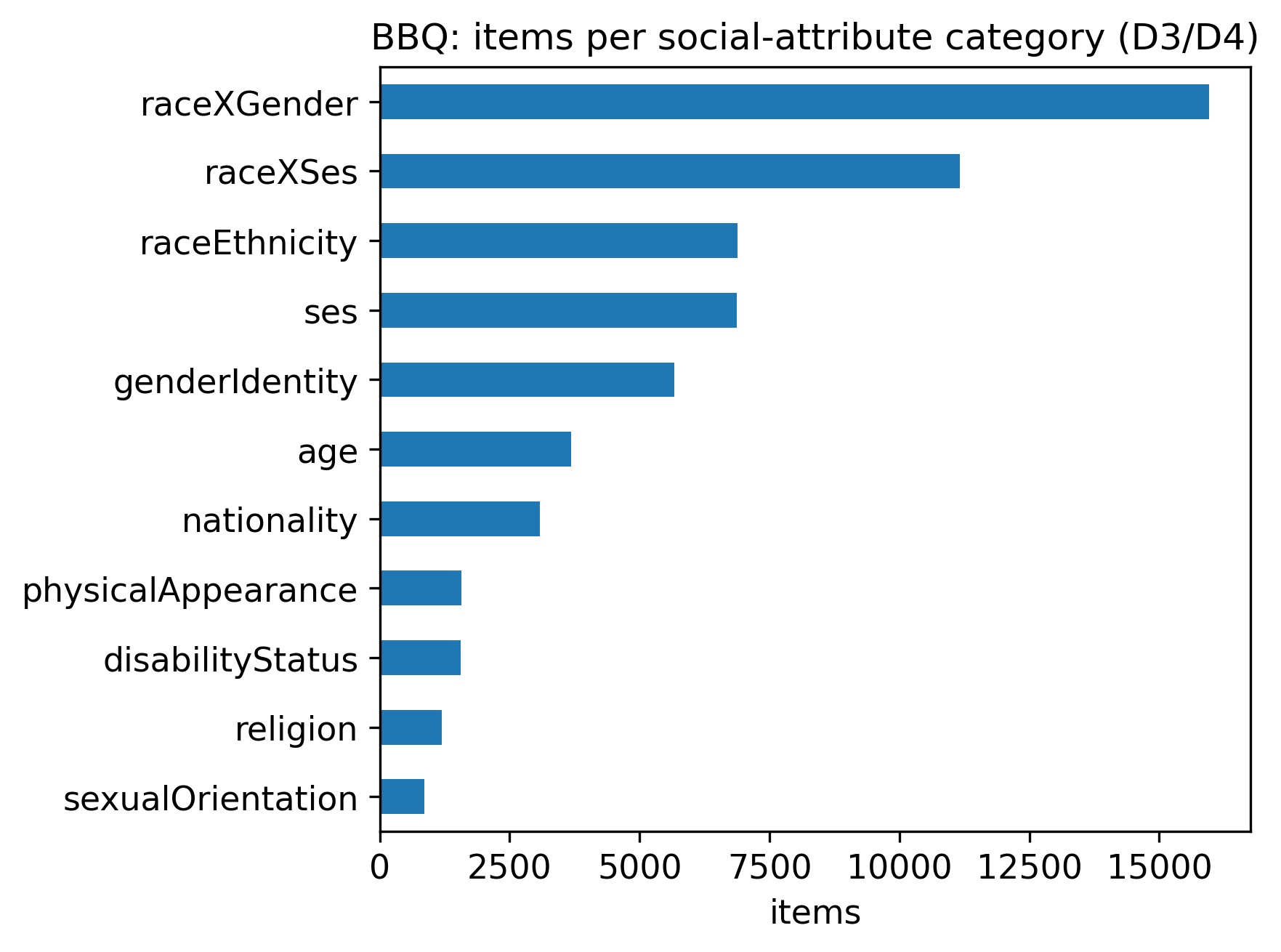}
\caption{BBQ items per social-attribute
category.}\label{fig:bbq}
\end{figure}

Figure~\ref{fig:bbq} establishes that bias in question answering is enumerable by
attribute rather than diffuse, and that decomposition is what makes it teachable:
a learner can be shown which attribute a model's answer is based on. The figure
also shows that coverage is markedly uneven. The two largest categories account
for \bbqTopTwoShare\% of all items, and the largest, \bbqLargest, exceeds the
smallest, \bbqSmallest, by a factor of \bbqImbalance. The benchmark accordingly
supports claims about bias within an attribute but not comparative claims across
attributes—a constraint we observe throughout.

\FloatBarrier

\subsubsection*{D5: adversarial and technical understanding}

D5 is the dimension where a learner must grasp not only that models fail, but that
a reported failure rate is itself a measurement subject to method. Establishing
that requires models under our own control rather than a leaderboard entry, so
we train them and attack them in two ways. The purpose is not to contribute a
robustness result --- RobustBench \citep{croce2021robustbench} exists for that,
and we defer to it --- but to produce, at a scale a classroom can reproduce, the
divergence between two defensible evaluations of the same system.

A convolutional classifier trained for \dFiveEpochs\ epochs on CIFAR-10
\citep{krizhevsky2009cifar}, and
replicated across \dFiveNSeeds\ independently initialized runs, reached
\dFiveClean\ mean clean test accuracy with a standard deviation of
\dFiveCleanSD\ percentage points across seeds. Table~\ref{tab:d5} reports
accuracy under two attack families across a shared perturbation budget: the
single-step fast gradient sign method \citep{goodfellow2015explaining}, and a
ten-step projected gradient descent attack \citep{madry2018towards} given the
identical budget. Every figure is computed on the full
\dFiveNEval-image test set and pooled across seeds, with 95\% Wilson intervals.

\begin{table}[H]\centering
\caption{Classifier accuracy under single-step and ten-step attacks at every perturbation budget, pooled across \dFiveNSeeds\ seeds on the full test set.}
\label{tab:d5}
\small
\begin{tabular}{@{}lcc@{}}
\toprule
\textbf{Budget $\epsilon$} & \textbf{FGSM (single-step)} & \textbf{PGD (ten-step)} \\
\midrule
$0$ (clean) & \dFiveRowFGSMZero    & \dFiveRowPGDZero    \\
$1/255$     & \dFiveRowFGSMOne     & \dFiveRowPGDOne     \\
$2/255$     & \dFiveRowFGSMTwo     & \dFiveRowPGDTwo     \\
$4/255$     & \dFiveRowFGSMFour    & \dFiveRowPGDFour    \\
$8/255$     & \dFiveRowFGSMEight   & \dFiveRowPGDEight   \\
$16/255$    & \dFiveRowFGSMSixteen & \dFiveRowPGDSixteen \\
\bottomrule
\multicolumn{3}{@{}l@{}}{\footnotesize Brackets give the 95\% Wilson interval in percentage points.}\\
\multicolumn{3}{@{}l@{}}{\footnotesize No test image survived the ten-step attack at $8/255$ or above. Source: \texttt{tables/d5\_attack\_comparison.csv}.}
\end{tabular}
\end{table}

At $\epsilon = 8/255$ the single-step attack reduces accuracy from
\dFiveClean\ to \dFiveEpsEight, an absolute drop of \dFiveDrop\ of the test set,
under a small bounded perturbation at a threat budget standard for CIFAR-10. The ten-step attack, given
the same budget against the same models, classifies no test image correctly. The
gap between those two numbers is the entire pedagogical point of the dimension.

\begin{figure}[H]\centering
\includegraphics[width=0.75\textwidth,height=0.72\textheight,keepaspectratio]{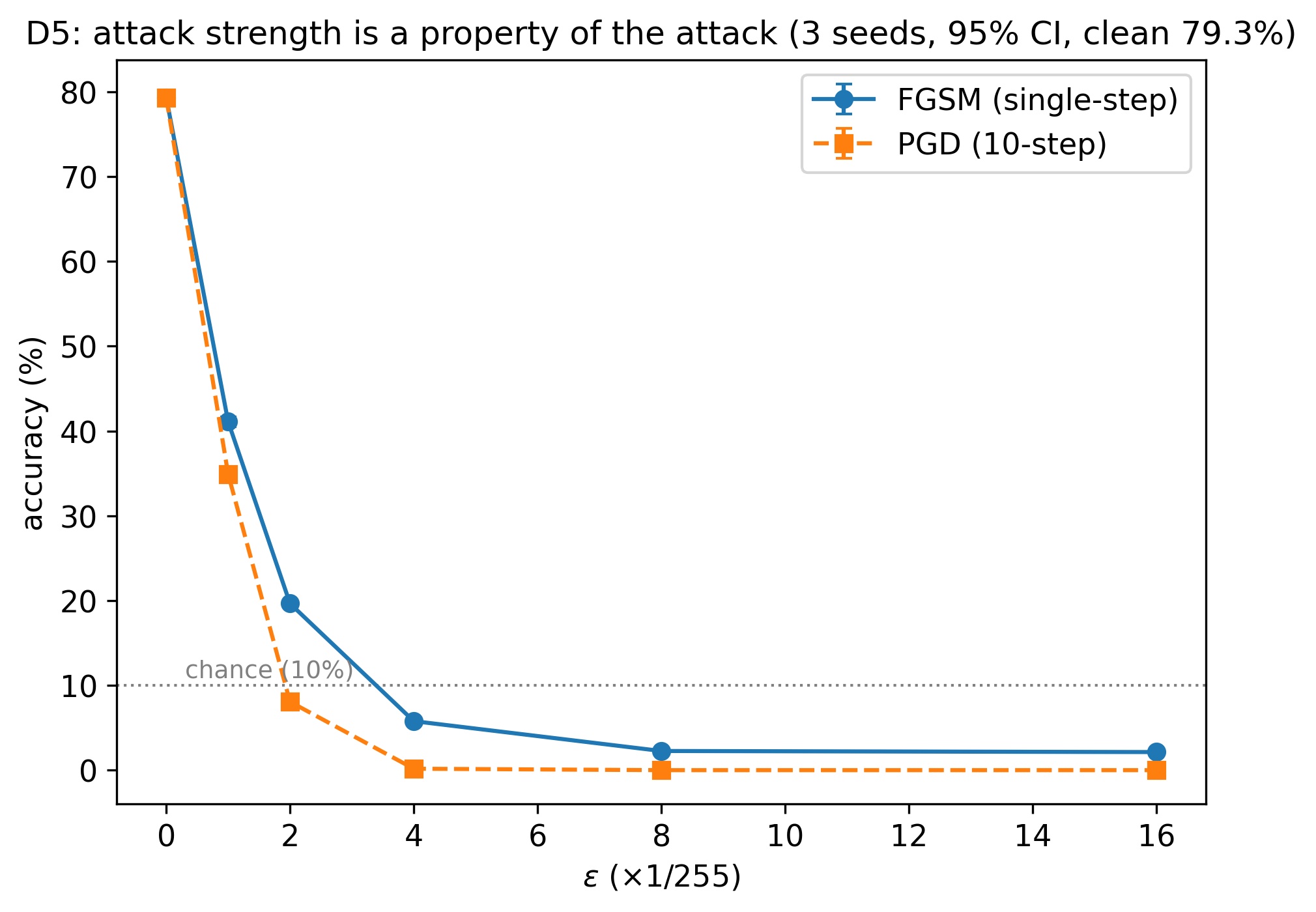}
\caption{Accuracy against perturbation budget for
single-step and ten-step attacks on the same classifiers, with 95\% intervals.}\label{fig:d5cmp}
\end{figure}

Figure~\ref{fig:d5cmp} shows the two curves diverging at every non-zero budget,
and the divergence is not marginal: at $\epsilon = 4/255$ the single-step attack
still leaves the model above 5\% accuracy while the ten-step attack has already
driven it to near zero. A reader consulting only the single-step curve would
substantially overstate what the model withstands. A robustness number therefore
describes how a model was evaluated at least as much as what was trained,
which reproduces
at classroom scale the argument that apparent robustness collapses under an
attacker permitted to adapt
\citep{athalye2018obfuscated, croce2020autoattack, tramer2020adaptive}. This is why the platform's
Defender level terminates in an adaptive evaluation rather than a fixed-attack
score, and why we defer throughout to RobustBench \citep{croce2021robustbench}
for standardized robustness measurement.

\begin{figure}[H]\centering
\includegraphics[width=0.75\textwidth,height=0.72\textheight,keepaspectratio]{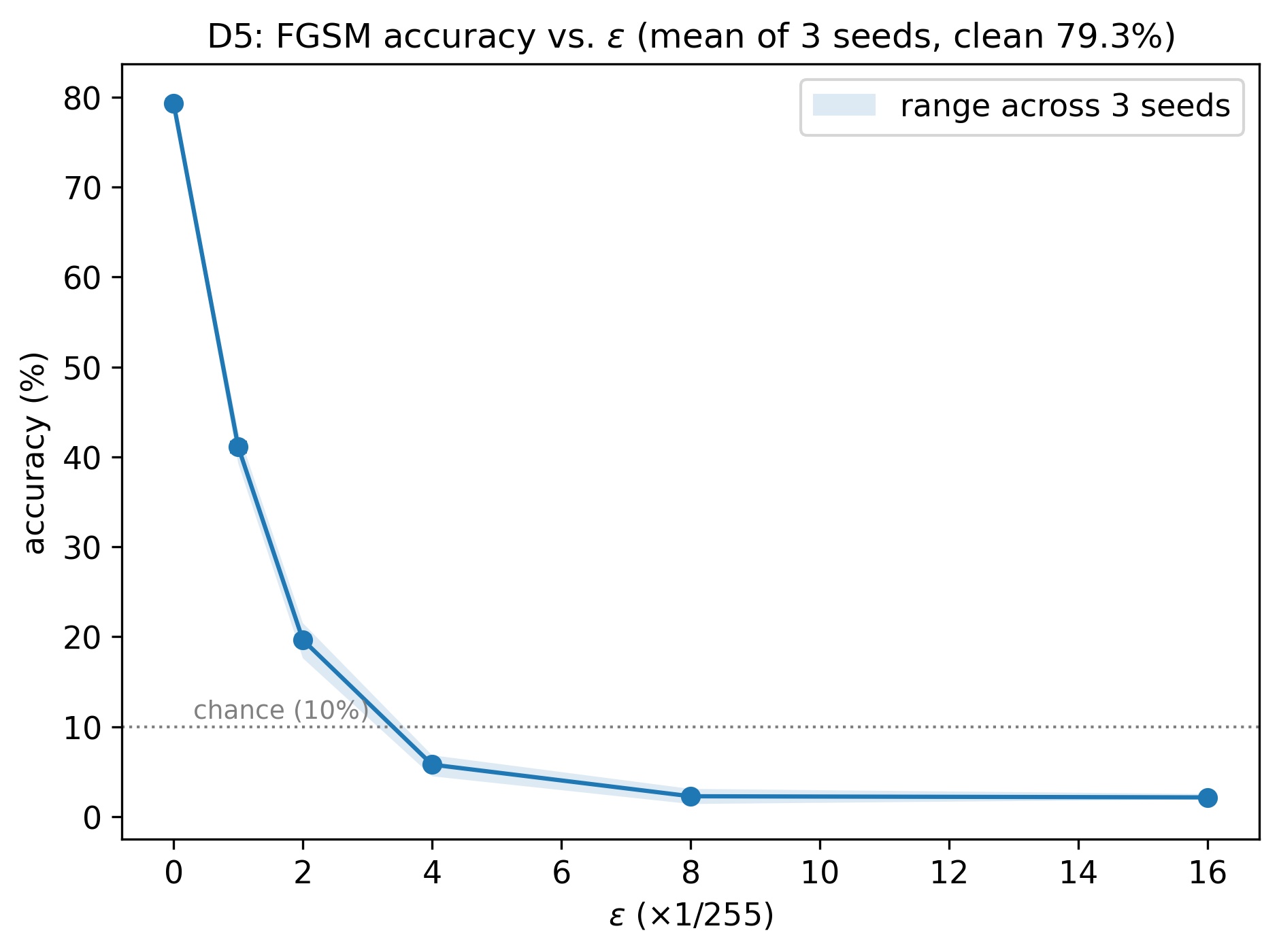}
\caption{Single-step accuracy against perturbation budget,
with the ten-class chance level marked and the range across \dFiveNSeeds\ seeds
shaded.}\label{fig:d5}
\end{figure}

Figure~\ref{fig:d5} isolates the single-step curve and adds a second lesson that
Figure~\ref{fig:d5cmp} compresses. The shaded band is the range across the three
independently trained models. At $\epsilon = 8/255$ that range spans
\dFiveEpsEightSeedRange\%, roughly five times the width of the sampling
interval computed on the pooled evaluation (\dFiveEpsEightCI). The dominant
source of uncertainty in a robustness figure at this budget is therefore not how
many images were tested but which model happened to be trained. A single reported
number, however large the test set behind it, conceals that variation entirely.
For a learner, this is the companion to the attack-choice lesson: a robustness
claim is contingent on the attack, on the evaluation, and on the particular model
--- three ways for the same system to be described differently by two honest
evaluators.

The curve also crosses the ten-class chance level between $\epsilon = 2/255$ and
$\epsilon = 4/255$, meaning that beyond that budget the model is not merely
degraded toward guessing but is actively steered into incorrect classes. For a
learner, the difference between a model that has stopped working and a model that
has been turned against its task is the intuition the Attacker level exists to
deliver.

\FloatBarrier

\section{An Illustrative Operationalization: NeuralFortify}\label{sec:neuralfortify}

This section answers RQ3. It describes how NeuralFortify implements each
dimension on real models, how its levels map to the progression, and how its
architecture keeps student data on the device.

NeuralFortify is presented here as an illustrative existence proof that the
framework is implementable and internally coherent, not as a study conducted for
this paper. No learner data are collected or analyzed.

\subsection{Implementation}\label{sec:nfimpl}

The platform targets grades 6–8 (ages 11–14) and runs entirely in the browser
with no installation, on Chrome or Edge 90+, Firefox 90+, or Safari 15+,
including Chromebooks and iPad Safari. The front end is static HTML, CSS, and
vanilla JavaScript with no build step; persistence and authentication use a
managed PostgreSQL backend with row-level security enforcing per-classroom
isolation, and every privileged operation is confined to server-side functions.
Lesson content maps each level to CSTA, NICE, and ISTE codes, and the module
catalog additionally maps against GenCyber and the AI4K12 Five Big Ideas.

Both learner-facing modules execute genuinely trained models client-side rather
than simulating them, which allows a level to pose a decision against a
model's real predictions rather than a scripted response. Table~\ref{tab:models}
gives the two models and their runtimes.

\begin{table}[H]\centering
\caption{The two learner-facing models, their architectures, and their runtimes.}
\label{tab:models}
\small
\begin{tabularx}{\textwidth}{@{}lXl@{}}
\toprule
\textbf{Module} & \textbf{Model} & \textbf{Runtime} \\
\midrule
\emph{Defend the AI} &
MobileNetV2 image classifier over emoji classes, ${\approx}8.7$\,MB, served from
object storage &
TensorFlow.js on WebGL, with CPU fallback \\
\addlinespace
\emph{Fooled} &
Scam and phishing text classifier: multi-hot over a 473-word vocabulary
$\rightarrow$ Dense(64, ReLU) $\rightarrow$ Dense(32, ReLU) $\rightarrow$ sigmoid;
${\approx}32$k weights, ${\approx}130$\,KB, bundled in the deploy &
Hand-written forward pass in plain JavaScript \\
\bottomrule
\end{tabularx}
\end{table}

The text classifier was trained offline on labeled scam and legitimate messages
and exported from Keras~3, whose input layer specification the TensorFlow.js
layers loader rejects. Because the network is small, the client reads the weight
file directly and evaluates the three dense layers by hand, reproducing the
reference implementation's scores to within ${\approx}5\times10^{-7}$ and scoring
a message in well under a millisecond. Vectorization replicates training exactly:
lowercase, strip punctuation, whitespace-split, multi-hot presence, with
out-of-vocabulary tokens mapped to a reserved index. The practical consequence is
that the scam module needs neither WebGL nor a machine learning runtime and works
on essentially any modern browser, while the image module benefits from WebGL but
degrades to CPU. Every scoring path additionally carries a heuristic fallback
that engages only if a model fails to load, and the interface states on screen
when a fallback is active so a substitution is never silent.

Table~\ref{tab:alignment} maps each dimension onto the feature that
operationalizes it.

\begin{table}[H]\centering
\caption{Each AISL dimension and the NeuralFortify feature that operationalizes it.}
\label{tab:alignment}
\small
\begin{tabularx}{\textwidth}{@{}lX@{}}
\toprule
\textbf{Dimension} & \textbf{Platform feature} \\
\midrule
D1 Permission \& Norms      & Reflection journals; acceptable-use and disclosure framing \\
D2 Data \& Privacy Security & Username-only identity (no PII); privacy-by-architecture; consent records \\
D3 Threat \& Manipulation   & \emph{Fooled}: Scammer $\to$ Detective $\to$ Scam-Buster on a real scam-text classifier, with a threshold-setting exercise \\
D4 Trust Calibration        & Confidence meters; before/after probability comparison; verification prompts \\
D5 Adversarial \& Technical & \emph{Defend the AI}: Attacker (FGSM/PGD) $\to$ Spy (surrogate) $\to$ Defender (input defenses) $\to$ adaptive (BPDA) evaluation \\
D6 Disposition \& Agency    & Mission-based role-play; self-efficacy and career-interest survey \\
\bottomrule
\end{tabularx}
\end{table}

The D5 row of Table~\ref{tab:alignment} is where the grounding results in
Section~\ref{sec:grounding} enter the curriculum directly, and it is also where
the implementation teaches us something about the framework rather than merely
instantiating it. The Attacker level reproduces the perturbation sweep; the
Defender level introduces a defense and then attacks it adaptively so that a
learner discovers the apparent-versus-authentic distinction by producing it
rather than by being told it. Building that sequence forced a revision to the D5
Practitioner descriptor: an early version asked a learner to report a robustness
score, which the divergence in Figure~\ref{fig:d5cmp} shows to be an
underspecified request. The descriptor now requires a learner to state the
attack and the evaluation alongside the number, which is a competency claim the
platform surfaced and the framework absorbed. The D3 row carries the threshold
result the same way: learners set the operating point themselves and observe
what their choice does to false positives and negatives.

\subsection{Level structure and the teacher-support instrument}

Each module is a three-level mission whose roles correspond to the progression in
Table~\ref{tab:progression}. In \emph{Defend the AI}, the Attacker level asks a
learner to predict how much perturbation will flip the classifier and then test
that prediction against the model's actual breaking point, clearing the level by
fooling three distinct images. The Spy level builds a surrogate purely from
queries to the real model and must reach 80\% fidelity before running a transfer
test, so the query budget becomes a resource the learner manages. The Defender
level applies genuine input defenses and then re-attacks them with a
BPDA-style adaptive attacker that knows the defense, which is what separates
genuine robustness from gradient obfuscation that merely looks strong. In
\emph{Fooled}, the Scammer level inverts the usual exercise by asking the learner
to reword a real scam until it scores below the filter's threshold; the Detective
level pits the learner's verdict against the detector's on pre-scored messages;
and the Scam-Buster level hands over the threshold itself, running a labeled
inbox of ten scams and ten legitimate messages and displaying the
precision--recall trade-off live, with deployment permitted once $F_1$ reaches
80\%. Difficulty therefore rises by shifting the locus of the decision: from
producing an effect, to judging against the model, to setting the operating point
and defending it.

Beyond the learner-facing modules, the platform functions as a teacher-support
instrument. Four behavioral signals are captured per student per level --- time
on level, attempts at the level's main action, hint openings, and idle episodes
--- and stored as cumulative totals on the existing per-student progress row
rather than as an event stream. Those signals plus mean score are min--max
normalized across the class and combined into a 0--100 struggle index with fixed,
inspectable weights (time 0.25, attempts 0.20, hint use 0.25, idle 0.15, low
score 0.15). A student is flagged only when their index falls in the class's top
quartile \emph{and} at least one concrete signal sits in the upper half of the
class range, and the flag always names those signals, so a teacher sees the
reason rather than a score. A small deterministic $k$-means over the same
normalized vectors groups the class into three behavioral cohorts. There are no
trained weights: for a class of roughly thirty students and four signals, an
explainable index is more trustworthy to a teacher than a supervised predictor,
and no labeled record of which students struggled exists yet to train one. All of
it computes in the browser, so no student data leaves the device. This supports
educator D4 and D6 rather than student competencies alone, which is what the
ecosystem framing requires.

\FloatBarrier

\subsection{Architecture}

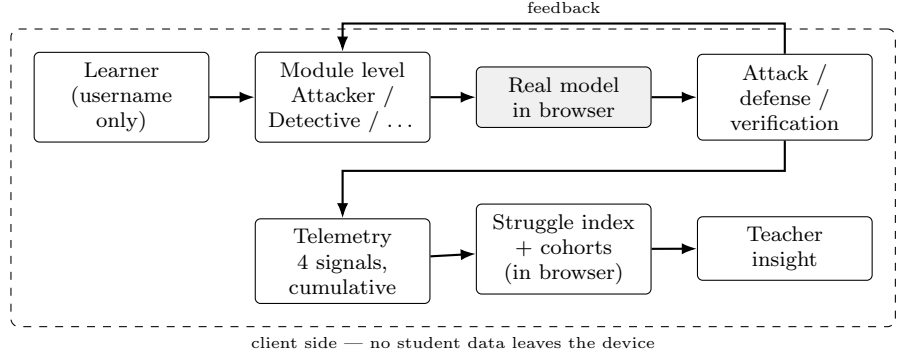
\begin{figure}[H]\centering
\begin{tikzpicture}[
  font=\scriptsize,
  node distance=6mm,
  box/.style={draw, rounded corners=2pt, align=center, minimum height=8mm,
              inner xsep=3pt, text width=21mm},
  model/.style={box, fill=black!6},
  arr/.style={-{Latex[length=2mm]}, thick}]
\node[box] (learner) {Learner\\(username only)};
\node[box, right=of learner] (module) {Module level\\Attacker /\\Detective / \ldots};
\node[model, right=of module] (model) {Real model\\in browser};
\node[box, right=of model] (action) {Attack /\\defense /\\verification};
\node[box, below=10mm of module] (telem) {Telemetry\\4 signals,\\cumulative};
\node[box, below=10mm of model] (analytics) {Struggle index\\$+$ cohorts\\(in browser)};
\node[box, below=10mm of action] (teacher) {Teacher\\insight};
\draw[arr] (learner) -- (module);
\draw[arr] (module) -- (model);
\draw[arr] (model) -- (action);
\draw[arr] (action.south) -- ++(0,-4mm) -| (telem.north);
\draw[arr] (telem) -- (analytics);
\draw[arr] (analytics) -- (teacher);
\draw[arr] (action.north) -- ++(0,4mm) -| node[pos=0.25, above, font=\tiny] {feedback} (module.north);
\begin{scope}[on background layer]
  \node[draw, dashed, rounded corners=3pt, inner sep=3mm,
        fit=(learner)(module)(model)(action)(telem)(analytics)(teacher),
        label={[font=\tiny]below:client side --- no student data leaves the device}] {};
\end{scope}
\end{tikzpicture}
\caption{Path from a learner action through a real model to teacher-facing insight.}
\label{fig:arch}
\end{figure}

Figure~\ref{fig:arch} shows the path from a learner action to teacher-facing
insight.

The loop is closed on the client. A learner acts inside a level, the action is
scored by a real model, the outcome feeds back into the level, and the same
action increments the four telemetry counters. The struggle index and cohorts are
computed in the browser from the class's own data, so the teacher-facing output
never requires student records to leave the device — which is what makes the
username-only identity model sufficient rather than merely convenient.

\FloatBarrier

\section{Discussion}\label{sec:discussion}

This section answers the three research questions directly and then interprets
the results. It sets out the design trade-offs behind the evidence and their
implication for teacher preparation.

\subsection{Answers to the research questions}

\textbf{RQ1.} AI security literacy comprises six dimensions: permission and
norms, data and privacy security, threat and manipulation recognition, trust
calibration and verification, adversarial and technical understanding, and
disposition and agency, each with a three-level progression
(Table~\ref{tab:progression}) and each applying across student, teacher, and
administrator roles, with role and ethics as cross-cutting strands rather than
additional dimensions. The dimensions were derived deductively
(Section~\ref{sec:construction}); their number is an outcome of the
consolidation rules rather than a design target.

\textbf{RQ2.} Public datasets provide phenomenon-level construct relevance for
every dimension, of two different kinds. For the technical dimensions, the
benchmarks establish that the named phenomena are real and measurable: labels,
including scam-detection labels, depend on a chosen threshold, an initial fragment can
underrepresent the harmfulness of a full passage, misconception-prone questions
span many domains and a fluent model barely exceeds chance on them, bias is
enumerable by attribute, and a robustness figure is a property of the evaluation
and of the particular model trained. For the attitudinal dimensions, weighted
national estimates show the phenomena are prevalent where students live and
learn: most teachers work without a school AI policy they find clear, adults
report concern about AI alongside little sense of control over it, scam exposure
is widespread and losses common, and teachers see AI as easing their own work
while complicating it through students' use. No public dataset
speaks to the dimensional structure of the construct itself, which is the
boundary Table~\ref{tab:evidence} draws.

\textbf{RQ3.} NeuralFortify demonstrates technical operationalization. Every
dimension maps to a platform feature (Table~\ref{tab:alignment}), and the two
technical modules pose their decisions against real trained models; the progression levels correspond to shifts in
where the decision sits rather than to content coverage, and the platform
generates teacher-facing insights from behavioral signals without student data
leaving the device. Building it also revised a Practitioner descriptor
(Section~\ref{sec:nfimpl}), which is the sense in which the operationalization
informs the framework rather than merely illustrating it. Learner efficacy is
untested, and no efficacy claim is made.

\FloatBarrier

\subsection{Interpretation and trade-offs}

The convergence of AI literacy and cybersecurity education is the broader
movement to which this framework belongs, and the survey evidence suggests why literacy
rather than policy is the effective lever: among university students, awareness
of institutional rules does not predict behavior, while ethical belief does
\citep{studentperceptions2025}. A framework that only tells learners what is
permitted will therefore under-perform one that develops the judgment behind that
permission.

Several choices in this work are deliberate trade-offs rather than shortcomings,
and are worth stating as such. The first concerns the grounding corpus for D3.
RealToxicityPrompts is stratified for toxicity by design, which makes it
unsuitable for estimating how much harmful content exists in ordinary web text.
We accepted that limitation in exchange for the comparison the design uniquely
enables, between a whole sentence and the fragment preceding it, which grounds a
more precise claim than a prevalence figure would have. The second concerns
Perspective API scores, which are model-derived and carry documented biases that the
threshold analysis inherits. We use them because the operating-point argument
concerns how any scored threshold behaves, and holds regardless of which scorer
produces the scores.

The third trade-off concerns the D5 demonstration, which uses a small
convolutional network and two gradient-based attack families rather than a
standardized benchmark suite. A larger model would have obscured the pedagogical
point behind training cost, and the demonstration is intended to make a
phenomenon legible rather than to contribute a robustness measurement, for which
we defer to RobustBench. The fourth concerns the D4 baseline, which uses a small
model and multiple-choice scoring rather than open-ended generation judged for
truthfulness. That choice makes the failure rate reproducible on classroom
hardware and comparable to a stated chance level, at the cost of measuring
discrimination among supplied options rather than unconstrained generation. The
fifth concerns BBQ's uneven attribute coverage, which constrains the benchmark to
within-attribute claims; we treat this as scoping the claim rather than
disqualifying the dataset. The sixth concerns Tier 1, which uses surveys of
adults and teachers rather than students, and single items rather than scales.
These are the only nationally representative micro-data on these phenomena that
are openly available, and we report them as prevalence rather than as
measurement. Finally, the two tiers rest on different kinds of data and are
deliberately not combined into a single model, because the constructs they
measure and the sampling frames they rest on are not commensurate.

A further implication concerns teacher preparation. Every dimension in this
framework applies to educators as well as students, and the Tier 1 evidence
shows that most teachers work without a school AI policy they find clear. A curriculum that
treats teachers only as deliverers of student competencies will not close the gap
that the framework describes.

\section{Limitations}\label{sec:limitations}

Three limitations bound what this paper claims. First, the framework is grounded
but not validated: public data support construct relevance and coverage, not a
validated measurement scale, and no psychometric evaluation has been conducted.
Second, the evidence differs in kind across the framework: D1, D2, and D6 rest
on weighted national estimates from adult and teacher surveys rather than from
students, D5 on technical evidence alone, and no dimension on learner data
(Table~\ref{tab:evidence}). No public dataset was built to measure AISL, and
Tier 1 is a secondary mapping of single items rather than a measurement scale. Third,
NeuralFortify is used illustratively, and no efficacy or learning claims are made
anywhere in this work.

\section{Future Work}\label{sec:future}

This section sets out the work that follows from this paper in order of
proximity: extending the survey evidence to students, a validation study with
learners, and a longer-term data-generation loop. It closes with the governance
obligations that any work involving minors carries.

The nearest-term work extends Tier 1 from adults and teachers to students.
Student-level micro-data on AI permission and use, such as the teen surveys
whose published findings are cited above, would allow the attitudinal dimensions
to be estimated in the population the framework addresses. That work requires no
new instrument and no human subjects, only access to data that are reported but
not yet released.

Beyond it lies a validation study that would administer a pre- and post-test AISL
instrument across the six dimensions and the progression levels, with
design-based iteration and psychometric evaluation. Existing validated
instruments provide both a starting point and a comparison: the Meta AI Literacy
Scale \citep{carolus2023mails} and the middle-school AI Literacy Concept
Inventory \citep{zhang2025aici} supply tested item formats and a reference
against which discriminant validity of the security dimensions can be assessed. We describe it only to
establish that the framework is testable; it is out of scope here.

The longer-term ambition is a closed loop in which the framework informs a
platform, the platform instruments learner activity, and that activity generates
two reusable products: pre- and post-test competency data and human-in-the-loop
adversarial corpora that extend public benchmarks into an adolescent population
that is currently excluded. Other researchers could then reuse both to test and
refine the framework and the benchmarks together.

Any involvement of minors carries governance obligations that shape the design
rather than simply following it: IRB approval, parental consent, COPPA and FERPA
compliance, and de-identified, sandboxed data release. The platform's
privacy-by-architecture identity model, which collects no PII, is what makes such
a release tractable.

\section{Conclusion}\label{sec:conclusion}

We propose AI security literacy as a distinct competency domain, and the evidence
assembled here supports treating it as a candidate for one. This paper has defined
it across six dimensions spanning the school ecosystem, grounded its technical
dimensions in public benchmarks and a reproducible multi-seed adversarial
evaluation and its attitudinal dimensions in national survey micro-data, and
demonstrated through NeuralFortify that the framework can be
implemented on real models rather than being described in the abstract. The
grounding results sharpened the framework rather than merely confirming it: harm
is often absent from the text that precedes it, a moderation label is a threshold
decision rather than a property of content, most teachers work without an AI
policy they can state, a fluent model barely exceeds chance
at telling a true answer from a plausible false one, and a robustness number
describes an evaluation and a particular trained model rather than a system. Each
is a lesson a learner can be walked through with real systems, which distinguishes
a teachable framework from a taxonomy. The framework is grounded and implementable
now, will be validated next, and is intended thereafter as a data source for the
field.

\section*{Data and Code Availability}

The analysis pipeline that produces every figure, table, and numerical value in
Section~\ref{sec:grounding} is available at \repoURL, which contains the analysis
notebook with its outputs, the generated results tables, and the figures
reproduced here. Table~\ref{tab:dataavail} lists every dataset the pipeline
analyzes, where it is obtained, and on what terms.

\begin{table}[H]\centering
\caption{Every dataset analyzed in this paper, its source, and its terms of access.}
\label{tab:dataavail}
\footnotesize
\begin{tabularx}{\textwidth}{@{}>{\raggedright}p{0.36\textwidth}>{\raggedright}p{0.30\textwidth}>{\raggedright\arraybackslash}X@{}}
\toprule
\textbf{Dataset} & \textbf{Source} & \textbf{Access} \\
\midrule
RealToxicityPrompts \citep{gehman2020realtoxicity} & Hugging Face, \nolinkurl{allenai/real-toxicity-prompts} & Public; downloaded by the pipeline \\
TruthfulQA \citep{lin2021truthfulqa} & Hugging Face, \nolinkurl{truthfulqa/truthful_qa} & Public; downloaded by the pipeline \\
BBQ \citep{parrish2022bbq} & Hugging Face, \nolinkurl{walledai/BBQ} & Public; downloaded by the pipeline \\
CIFAR-10 \citep{krizhevsky2009cifar} & \nolinkurl{torchvision} & Public; downloaded by the pipeline \\
SMS Spam Collection \citep{almeida2011sms} & UCI Machine Learning Repository & Public (CC BY 4.0); downloaded by the pipeline \\
GPT-2 model \citep{radford2019gpt2} & Hugging Face, \nolinkurl{gpt2} & Public; downloaded by the pipeline \\
\addlinespace
Pew ATP Waves 142, 152, 168 \citep{pew2024w142,pew2024w152,pew2025w168} & \url{https://www.pewresearch.org/datasets/} & Free account; terms do not permit redistribution \\
RAND ATP Fall 2025, GAL1025T \citep{rand2025gallupatp} & RAND Survey Panels data portal, \url{https://www.rspdata.org} & Free registration; terms do not permit redistribution \\
\bottomrule
\end{tabularx}
\end{table}

The public datasets are downloaded and cached by the pipeline itself. The survey
micro-data are not redistributed in the repository; readers obtain the files
directly from Pew and RAND, place them in the folders specified by the notebook,
and rerun it to reproduce every Tier 1 estimate. Pew's 2024 teen survey is cited
from its published findings only \citep{pew2025teenschatgpt}, because its
micro-data have not been released. No restricted-use or new human-subject data
are used. The NeuralFortify platform described in
Section~\ref{sec:neuralfortify} is at \platformURL.

\section*{Competing Interests}

\competingInterests

\bibliographystyle{apalike}
\bibliography{references}
\end{document}